\documentclass[10pt]{bmc_article}

\usepackage{amsmath}
\usepackage{amsfonts}
\usepackage{graphicx}
\usepackage{latexsym}
\usepackage[square,numbers,compress]{natbib}
\usepackage{url}  
\usepackage{ifthen}  
\usepackage{multicol}   
\usepackage[latin1]{inputenc} 
\usepackage[colorinlistoftodos]{todonotes} 

\DeclareMathOperator*{\Mn}{\ifmmode{{\bf M(n;\R) }}\else{$\bf M(n;\R)  $}\fi}

\usepackage[colorinlistoftodos]{todonotes}

\graphicspath{{../figures/}}

\newboolean{publ}

\newenvironment{bmcformat}{\begin{raggedright}\baselineskip20pt\sloppy\setboolean{publ}{false}}{\end{raggedright}\baselineskip20pt\sloppy}

\begin{document}
\begin{bmcformat}


\title{Phylogenetic quantification of intra-tumour heterogeneity}
 

\author{Roland F Schwarz\correspondingauthor$^{1,2,3}$%
  \email{Roland F Schwarz - rfs32@cam.ac.uk}%
  \and
  Anne Trinh$^{1,2}$%
  \email{Anne Trinh - anne.trinh@cruk.cam.ac.uk}%
  \and
  Botond Sipos$^{3}$%
  \email{Botond Sipos - sbotond@ebi.ac.uk}%
  \and
  James D Brenton$^{1,2,4}$%
  \email{James Brenton - james.brenton@cruk.cam.ac.uk}%
  \and
  Nick Goldman$^{3}$%
  \email{Nick Goldman - goldman@ebi.ac.uk}%
  \and
  Florian Markowetz\correspondingauthor$^{1,2}$%
  \email{Florian Markowetz - florian.markowetz@cruk.cam.ac.uk}%
}


\address{%
    \iid(1)University of Cambridge, Cambridge UK\\
    \iid(2)Cancer Research UK Cambridge Institute, Cambridge, UK\\
    \iid(3)European Bioinformatics Institute, Hinxton, UK\\
   \iid(4)Department of Oncology, University of Cambridge, UK
}%

\maketitle


\vspace{-5mm}
\begin{abstract}
  \paragraph*{Background:} Intra-tumour heterogeneity (ITH) is the
  result of ongoing evolutionary change within each cancer. The
  expansion of genetically distinct sub-clonal populations may explain
  the emergence of drug resistance and if so would have prognostic and
  predictive utility. However, methods for objectively quantifying ITH
  have been missing and are particularly difficult to establish in
  cancers where predominant copy number variation prevents
  accurate phylogenetic reconstruction owing to horizontal
  dependencies caused by long and cascading genomic rearrangements.
      
  \paragraph*{Results:} To address these challenges we present MEDICC, a
  method for phylogenetic reconstruction and ITH quantification based
  on a \textbf{M}inimum \textbf{E}vent \textbf{D}istance for
  \textbf{I}ntra-tumour \textbf{C}opynumber
  \textbf{C}omparisons. Using a transducer-based pairwise comparison
  function we determine optimal phasing of major and minor alleles,
 as well as evolutionary distances between samples, and are able to reconstruct
  ancestral genomes. Rigorous simulations and an extensive clinical study show
  the power of our method, which outperforms state-of-the-art
  competitors in reconstruction accuracy and additionally allows
  unbiased numerical quantification of ITH.

  \paragraph*{Conclusions:}  Accurate quantification and evolutionary inference are
  essential to understand the functional consequences
  of ITH. The MEDICC algorithms are
  independent of the experimental techniques used and are applicable to
  both next-generation sequencing and array CGH data.
\end{abstract}

\ifthenelse{\boolean{publ}}{\begin{multicols}{2}}{}


\section*{Background}
The study of intra-tumour genetic heterogeneity (ITH) is now a major
focus of cancer genomics research
\cite{Khalique2007,Khalique2009,Cooke2010, Navin2010, Shah2009,
  Campbell2010, Marusyk2012, Navin2011, Vermaat2012, Wu2012,
  Nik-Zainal2012, Gerlinger2012} due to its potential to provide
prognostic information \citep{Cooke2011a, Maley2006, Park2010} and to
explain mechanisms of drug resistance \citep{Cooke2011, Ding2010,
  Ding2012, Cowin2012}.
Quantifying heterogeneity and understanding its aetiology crucially
depends on our ability to accurately reconstruct the evolutionary
history of cancer cells within each patient. In many cancers, such as
high-grade serous ovarian cancer (HGSOC), most of this heterogeneity
is not reflected in point mutations but in genomic rearrangements and
endoreduplications that lead to aberrant copy number (CN) profiles
\citep{TCGA2011, Ng2012}. In these cases tree inference is hindered by
unknown phasing of parental alleles, horizontal dependencies between
adjacent genomic loci and the lack of curated CN profile databases to
use as a reference for probabilistic inference.  Therefore
heterogeneity and evolutionary divergence are typically quantified
using ad-hoc thresholds \citep{Cowin2012} and tree inference is often
done subjectively \citep{Nik-Zainal2012}. Approaches developed to
address this problem include Greenman et al. \citep{Greenman2012}, a
graph theoretical approach on signed reversals to order rearrangement
events, but this requires detailed annotation of
rearrangements in the data that may not be available, and the
algorithm does not generally infer global trees representing cancer
evolution within a patient. The \emph{TuMult} algorithm
\citep{Letouze2010} deals with underlying computational complexity by
considering only breakpoints --- locations on the genome where the CN
changes --- and by using total CN without phasing of parental alleles.
While simplifying the computational problem, this approach discards
potentially informative data.

Our aim is to establish numerical quantification of ITH per patient
from CN profiles that can routinely be acquired from clinical samples.
To this end, we have developed MEDICC (Minimum Event Distance for
Intra-tumour Copy-number Comparisons), a method for accurate inference
of phylogenetic trees from unsigned integer CN profiles. MEDICC
specifically addresses the following challenges associated with
CN-based phylogeny estimation. i) It makes use of the full CN
information across both parental alleles by estimating an optimal
\emph{phasing} of CN variants using the evolutionary information. ii)
It deals with \emph{horizontal dependencies} between adjacent genomic
loci and with multiple overlapping events by using efficient
heuristics. It can therefore work on complete genomic profiles instead
of breakpoints which allows the reconstruction of ancestral
genomes. iii) It implements statistical tests for evolutionary rates,
tests for phylogenetic structure and tests for the relationship
between clonal subpopulations to provide informative \emph{summary
statistics} for the reconstructed evolutionary histories and ITH.

We developed MEDICC and successfully applied it to the analysis of a
novel dataset of 170 CN profiles of patients undergoing neo-adjuvant
chemotherapy for HGSOC as described in our accompanying clinical study
(Schwarz \textit{et al.}  2013 \emph{in prep.}). In the following we
first give a more detailed description of the data and problems that
MEDICC addresses. We then introduce the MEDICC modelling framework
that guides all steps of the algorithm and which is then explained in
detail. We finish with a demonstration of MEDICC on a real-world
example of a case of endometrioid cancer and give simulation results
that compare it to competing methods.

\section*{Results and Discussion}
MEDICC was designed to work on human integer CN profiles that can
routinely be obtained from single nucleotide polymorphism (SNP) arrays
\citep{Greenman2010} or paired-end sequencing
\citep{Korbel2007,Yoon2009}. In both cases DNA content is quantified
relative to a diploid normal in windows along the genome. Point
mutations help to distinguish the two parental alleles and the
resulting profile comprises two vectors of integer CNs, representing
the absolute number of copies of that particular genomic segment in
both alleles; however the \emph{phasing} of those CNs to the two
parental alleles is unknown, unless two SNPs happen to be on the same
read or other external linkage information is available
\citep{Nik-Zainal2012}. Consequently, by definition (and for each
genomic segment independently), the larger of the two CNs is termed
the \emph{major} and the other the \emph{minor} CN, without any
information about which CNs belong together on the same allele
(Figure~\ref{fig:3steps} left). Without resolving this ambiguity,
tracing of individual events is not possible.

Another challenge is horizontal dependency in CN profiles. In contrast
to nucleotide substitution models where sites in a sequence are
modelled as independent and identically distributed (iid)
\citep{Felsenstein2003}, CN events often overlap and range across many
adjacent genomic regions and thereby introduce horizontal dependencies
that influence estimation of evolutionary divergence.

Given multiple such evolutionarily-related CN profiles, for example
from distinct primary and metastatic sites of the same patient,
phylogenetic inference in MEDICC then involves three steps: (i)
allele-specific assignment of major and minor CNs, (ii) estimation of
evolutionary distances between samples followed by tree inference and
(iii) reconstruction of ancestral genomes (Figure \ref{fig:3steps}).
All three steps are guided by a minimum evolution criterion. Similar
to early edit-distances for sequence analysis \citep{Levenshtein1966},
MEDICC counts the number of genomic events needed to transform one CN
profile into another and searches for the tree that minimises this
criterion.

\subsection*{MEDICC reconstructs evolutionary histories via  a minimum
  evolution criterion}
We model the evolution of genomic rearrangements through the following
set of events that have an observable effect on CN profiles: terminal
and interstitial deletions, as well as unbalanced translocations, are
single deletion events; tandem and inverted duplications are single
amplification events; and breakage fusion bridges are dual events
involving a duplication and a deletion \citep{Greenman2012}. We use a
finite-state automaton representation of genomic profiles and
finite-state transducers \citep{Mohri2003} for modelling and efficient
computing of the minimum-event distance based on these genomic events
(Figure \ref{fig:fst}A). Before going through the three steps of the
reconstruction process in detail it is necessary to introduce some
terminology; for a more thorough introduction into transducer theory
see \citep{Mohri2003, Mohri2004, Droste2009} and references therein.

\subsubsection*{The MEDICC modelling framework}
MEDICC models diploid genomic CN profiles as sequences over the
alphabet $\Sigma = \{0,\ldots,K,``X"\}$, where $\{0,\ldots,K\}$
represent integer CNs ($K$ is the maximum haploid CN) and $``X"$ is a
special character that separates the two alleles on which events can
happen independently. For example, the profile $1123002X0122002$
represents a chromosome with 7 regions distinguished, with the first
region present in one copy on one allele and absent in the other
allele; the second region present in one copy on each allele; and so
on up to the seventh region present in two copies on each allele. This
means that MEDICC deals with a maximum total CN of $2K$ in a diploid
genome. By default $2K=8$ which is the upper end of the dynamic range
of SNP arrays, but the alphabet can be extended easily without
changing the implementation. In this manuscript the terms ``sequence''
and ``(CN) profile'' are used interchangeably. 

CN profiles are
implemented as acceptors, weighted finite-state automata (FSA) that
can contain a single or multiple such profiles. The minimum-event
distance is implemented as a weighted finite-state transducer (FST,
\citep{Mohri2003}). FSTs are an extension of FSAs with input and
output symbols --- like pair-HMMs they emit or accept two sequences
simultaneously, meaning they model the events transforming on sequence
into another. Both FSAs and FSTs can be equipped with weights from a
semiring, enabling calculations to be weighted according to some
importance criterion. One of the most common semirings is the real
semiring (e.g. the weights represent probabilities), where weights are
multiplied along a path in the automaton and the total weight of a
sequence (or pair of sequences) is the sum (total probability) over
all possible paths generating that sequence. Equally popular is the
tropical semiring, also known as the Viterbi path, where weights are
summed along a path and the total weight is the minimum across all
those paths. In this case weights are often ``penalties'' or negative
log-probabilities for taking a certain path, similar to classical
pairwise sequence alignment in which mismatches and indels are
penalised with additive fixed scores. 

MEDICC uses the tropical
semiring for computing the minimum event distance, but the modularity
of the framework allows us to smoothly transition to probabilities at
a later stage by switching semirings without changing the algorithm.
In this tropical semiring a FST $T_1$ then assigns a score to two
sequences (represented as acceptors) $x$ and $z$ via

\begin{equation*}
  T_1[x,z] = \min_{p \in P} \sum_i w(p, i).
\end{equation*}

where $P$ is the set of all possible paths through the FST in which
the input and output symbols match with the sequences $x$ and $z$ and
$w(p,i)$ is the weight of that path at position $i$ in the
sequence. No score is returned for a pair of sequences for which no
valid path in $T_1$ exists. This leads to the definition of the
minimum-event distance, which governs all three steps of the
reconstruction process.

\subsubsection*{Constructing the minimum-event distance for CN profiles}
Figure \ref{fig:fst}B shows the one-step transducer $T_1$ that we use
to model single amplifications and deletions of arbitrary length and
that counts one event each time the amplification or deletion state is
entered. This is analogous to an affine gap cost model in classical
sequence alignment \citep{Durbin1998}. $T_1[x,z]$ therefore assigns to
each pair of sequences $(x,z)$ the minimum number of events necessary
to transform one sequence into another. At this point, however, not
all possible CN scenarios have a valid path (e.g. one event can
amplify ``1'' to ``2'' but not ``1'' to ``3''). To include all
possible changes across multiple events, $T_1$ is composed $K$ times
with itself \citep{Mohri2004}. In essence, composition describes the
chaining of FSTs, where the total weight of the composed transducer is
the total minimum score from the input sequence $x$ via intermediate
sequences $y_i$ to the target sequence $z$:

\begin{eqnarray*}
  T[x,z] &=& (T_1 \circ ... \circ T_1)[x,z]\\
  &=& \min_{y_1,\ldots, y_{K-1}} \left(T_1[x,y_1] + \ldots + T_1[y_{K-1}, z]\right)
\end{eqnarray*}

This gives rise to the FST $T$ that strictly adheres the modelled
biological constraints such as no amplification from zero. We call $T$
the \emph{tree} transducer: these biological constraints give it a
direction, and it is not guaranteed to return a distance for any pair
of CN profiles.

As we are interested in the minimum evolutionary distance between any
two sequences $x$ and $z$ via their last common ancestor (LCA) $y$,
the final distance FST $D$ is formed by composing $T$ with its inverse
(Figure \ref{fig:fst}C, \citep{Schwarz2010}), such that $D$ computes
the distance from a leaf node to the LCA ($T^{-1}$) and back ($T$) to
the other leaf node:

\begin{eqnarray*}
  D[x,z] & = & (T^{-1} \circ T)[x,z]\\
  & = & \min_y\left(T^{-1}[x,y] + T[y,z]\right)
\end{eqnarray*}

In the real semiring, and equipped with probabilities, this would be
analogous to classical phylogenetic reconstructions where a reversible
model of sequence evolution is used to compute the likelihood of the
subtree containing sequences $x$ and $z$ as the products of the
individual likelihoods of seeing $x$ and $z$ given their ancestor $y$
and summing over all $y$ \citep{Felsenstein1981}. In our case, $D$
equivalently computes the minimum number of events from $x$ to $z$ via
their LCA. This distance is symmetric and is guaranteed to yield a
valid distance for any pair of sequences. In the rest of the paper,
``distance'' refers to this minimum-event distance, unless stated
otherwise.

MEDICC therefore computes an evolutionary distance between two genomes
based on a minimum evolution criterion via their closest possible
LCA. Due to composition of the tree transducer $T$ with its inverse,
the resulting distance $D$ is a dissimilarity score that represents
(the logarithm of) the shortest-path approximation to a
positive-semidefinite kernel score \citep{Cortes2004,
  Schwarz2010}. This means that computing the evolutionary distances
between samples automatically places these samples in a
high-dimensional evolutionary space, where in addition to distances we
gain information about their relative position and angles. We term
this space the mutational landscape, on which we can directly apply
explorative analyses like PCA, classification with support-vector
machines and other machine learning techniques
\citep{Shawe-Taylor2004}. We use \emph{OpenFST}, an efficient
implementation of transducer algorithms \citep{Allauzen2007} to
achieve exact distance computation in quadratic time.

Following the minimum evolution principle, the overall objective is to
find a tree topology including ancestral states that minimises the
total tree length, i.e. the total number of genomic events along the
tree.  In the following we will describe how MEDICC achieves this in
its three step process.

\subsubsection*{Step 1: Evolutionary phasing of major and minor CNs}
As CN-changing events can independently occur on either or both of the
parental alleles, the allele-specific assignment, or phasing, of major
and minor CNs heavily influences the minimum tree length objective. We
use the evolutionary information between samples to solve these
ambiguities. Using our distance measure we can choose an optimal
phasing between a pair of diploid profiles that minimises the pairwise
distance between them (Figure \ref{fig:cfg}A). This respects the
distinct evolutionary histories of both alleles and finds a phasing
scenario in which the evolutionary trajectories between both haploid
pairs are minimal. To achieve this, a diploid profile is represented
as a single sequence in which the allele boundaries are marked by a
separator character as described above. From each pair of major
and minor input sequences we can generate up to $2^{L}$ such
concatenated sequences, where $L$ is the length of the input profile
(both alleles have the same length). This number is too large to
enumerate exhaustively, so in order to achieve a compact
representation of diploid profiles we make use of a context-free
grammar (CFG). Our implementation is related to the use of CFGs to
model RNA structures, where paired residues in stem regions are not
independent \citep{Durbin1998}. 

In our CN scenario a CFG represents different allele phasing
choices (see Figure~\ref{fig:cfg}B right). At every position in the diploid profile
we have a choice of using the major as the first allele and the minor
as the second or (``$|$'') vice versa (Figure \ref{fig:cfg}B
left). Each possible parse tree of the CFG then corresponds to one
phasing scenario out of the $2^{L}$ possibilities. When the distance
FST reads the separator it is forced to return to the match state
(initial state), thus guaranteeing that the total distance to another
profile equals the sum of the distances of the two alleles with no
events spanning different alleles. We represent CFGs algorithmically
by pushdown-automata in the FST library \citep{Allauzen2012}.

While this approach works well for finding phasing scenarios that
minimise the distance between one pair of profiles, we aim to find
phasing scenarios that jointly minimise the distances between all
profiles in the dataset. To reduce the computational complexity of
this task it is necessary to employ a heuristic. MEDICC searches for
the single profile that has minimum sum of distances to all sample
profiles, that is, the geometric median, through an iterative
search. This centre profile is then compared again to each individual
profile and the shortest path algorithm yields the choice of phasing that
minimises the distance between each profile and the centre. While this
is not guaranteed to return a globally optimal phasing scenario, it
has proven to perform very well in practise.

\subsubsection*{Step 2: Distances and tree reconstruction}
Once the alleles have been phased, pairwise evolutionary distances
between samples can be computed as the sum of the pairwise distances
between both alleles. MEDICC then uses the Fitch-Margoliash algorithm
\citep{Fitch1967} for tree inference from a distance matrix with or
without clock assumption. A test of clock-like events, available using
functionality in the accompanying R package \emph{MEDICCquant}, allows
us to determine which tree reconstruction algorithm is most
appropriate (see the section on quantification of ITH).

\subsubsection*{Step 3: Ancestral reconstruction and branch lengths}
From this point on we keep the topology of the tree fixed, and
traverse from its leaves to the root to infer ancestral CN profiles
and branch lengths. Ancestral reconstruction is possible because
cancer trees are naturally rooted by the diploid normal from which the
disease evolved. Reconstructing ancestral genomes allows us to
investigate e.g. the genomic makeup of the cancer precursor, the LCA
of all cancer samples in the patient. Events that across patients
frequently occur between the root of the tree and the precursor are
likely driver events of cancer progression. Ancestral reconstruction
also determines the final branch lengths of the tree. MEDICC infers
ancestral genomes for each allele independently using a variant of
Felsenstein's Pruning algorithm \citep{Felsenstein2003}. In
traditional ancestral reconstruction the total score
(likelihood/parsimony score) of the tree is computed in a downward
pass towards the root and ancestral states are then fixed in a second
upward pass, successively choosing the most likely/most parsimonious
states. 

In our scenario, the algorithm begins by composing each of the
$n$ terminal nodes with the tree transducer $T$, which yields $n$
acceptors holding all sequences reachable from that terminal node and
their respective distances. When the first two terminal nodes are
joined in their LCA the corresponding acceptors are intersected. The
resulting acceptor contains only those profiles that were contained in
both input acceptors and their corresponding weights are set equal to
the sum of the weights of the profiles in the input acceptors. In a
probabilistic framework the resulting acceptor is equivalent to the
conditional probability distribution $P(\text{subtree}
(x,z)~|~\text{LCA}~y) = P(x|y) P(z|y)$ for each possible LCA, where
the sum of distances again is replaced by the product of the
conditional probabilities of seeing a leaf node given its
ancestor. This intersection will still contain the vast majority of
all possible profiles, but each with a different total distance, and
without those that are prohibited by biological constraints. For
example, the ancestor cannot have a CN of zero at a position where at
least one of its leaf nodes has CN $>0$, as amplifications from zero
are not allowed.

Because after phasing each leaf node is represented by an acceptor
containing exactly one diploid sequence, computing this set of
possible ancestors is computationally feasible. However, because
during tree traversal we need to compose these sets of possible
profiles repeatedly with the tree transducer $T$, the result would
increase in size exponentially because it has to account for all
possible events of arbitrary length at each position in all
sequences. Therefore during tree traversal, when two internal nodes
have to be joined in their LCA, MEDICC reduces them to a single
sequence by choosing those two sequences with smallest distance to
each other. This fixes the profiles for those two internal nodes. This
procedure is continued until all internal nodes are resolved. Once all
ancestral CN profiles have been reconstructed the final branch lengths
are simply the distances between the nodes defining that branch in the
tree.

\subsection*{MEDICC improves phylogenetic reconstruction accuracy}
We assessed reconstruction accuracy using simulated data generated by
the \emph{SimCopy} R package \citep{Sipos2013} (see Methods). Random
coalescent trees were generated with \emph{APE}
\citep{Paradis2004}. To create an unbiased simulation scenario, genome
evolution was simulated using increasing evolutionary rates on the
sequence level using five basic genomic rearrangement events:
deletion, duplication, inverted duplication, inversion and
translocation (for details see Methods). Once the simulations were
complete, CNs were counted for each genomic segment and these CN
profiles were used for tree inference using the following three
methods: i) BioNJ \citep{Gascuel1997} tree reconstruction on a matrix
of euclidean distances computed directly on the CNs, ii)
breakpoint-based tree-inference using the \emph{TuMult} software
\citep{Letouze2010} and iii) MEDICC.  \emph{TuMult} additionally
requires array log-intensities as input. In order to keep the
comparisons unbiased, noiseless log ratios simulating CGH array
intensities for \emph{TuMult} were directly computed from the CN profiles. To
assess the relative abilities of the methods to correctly recover the
evolutionary relationships of the simulated CN profiles,
reconstruction accuracy was measured in quartet distance
\citep{Mailund2004} between the true and the reconstructed
tree. Quartet distance was chosen as it only considers topological
differences; branch lengths have widely different meanings in the
methods tested and as such are not comparable.

This simulation strategy is based on basic biological principles,
independent of the methods compared and \emph{a priori} does not
favour any of them. All simulations were repeated to cover a wide
parameter range, yielding qualitatively similar results.

The simulation results clearly show the improvement in reconstruction
accuracy of MEDICC over naive approaches (BioNJ on Euclidean
distances) and competing methods (TuMult) (Figure \ref{fig:sim}A). In
general, reconstruction accuracies increase with increasing
evolutionary rates. Especially when the amount of phylogenetic
information is limited, MEDICC outperforms other methods by a
significant margin. This may be because of two reasons:
firstly, in contrast to other methods MEDICC is capable of phasing the
parental alleles, thereby making much more effective use of the
phylogenetic information compared to methods that work on total CN
alone. Secondly, due to efficient and accurate heuristics, MEDICC can
deal with the horizontal dependencies imposed by overlapping genomic
events of arbitrary size and accurately computes distances between
them.

\subsection*{Evolutionary comparisons with MEDICC allow quantification of
  tumour heterogeneity}
As described earlier, the matrix of pairwise distances inferred by
MEDICC is the logarithm of a rational kernel matrix \citep{Cortes2004,
  Schwarz2010} which maps samples to a high-dimensional mutational
landscape. We reduce the dimensionality of this landscape through
kernel principal components analysis \citep{Scholkopf1998} where we
can use spatial statistics to derive numerical measures of ITH for
each patient.

Intra-tumour heterogeneity is a loose concept that describes the
amount of genomic difference between multiple cells or samples of the
same tumour. Two types of heterogeneity often of interest are
\emph{spatial} and \emph{temporal} heterogeneity. For example, spatial
differences might be observed from separate biopsies of a primary cancer
and a distant metastasis. Other changes may occur between different
time points, for example before and after chemotherapy. Average
distances between subsets of samples might be computed by any method
that returns dissimilarities between samples by simple
averaging. However, as clinical datasets are noisy, more robust
measures of distances between aggregated subsets of samples are
desirable.
\subsubsection*{Temporal heterogeneity}
We define temporal heterogeneity as the evolutionary distance between
the average genomic profiles between any two time points (e.g. at
biopsy before chemotherapy and at surgery after chemotherapy in the
case of neo-adjuvant treatment). In the mutational landscape (see
above) we are able to directly compute the centre of mass of a set of
genomic profiles (which would not be possible by working with
distances alone). We can then define temporal heterogeneity as the
distance between the centres of mass of the samples from those two time
points (Figure \ref{fig:ith}D). The advantage of this approach is that
we can use robust measures of the centre of mass (e.g. ignoring the
single most distant point) to estimate temporal heterogeneity. It
should be noted that this general approach can be used for determining
distances between any partitions of the samples in the dataset, for
example between groups of samples taken from different organs as a
measure of spatial heterogeneity.

\subsubsection*{The clonal expansion index}
Other complex aspects of ITH that cannot be easily derived from
distances alone include the ability of a tumour to undergo clonal
expansions \citep{Cooke2011}. The model here is that if the majority
of cancer cells are subject to strong selection pressure, such as from
chemotherapy, minor subclones with a distinctive selective advantage
may repopulate. This subpopulation would be expected to coalesce early
and will show a greater than expected divergence (relative to neutral
evolution) from other remaining clones. This model is similar to
analyses of clonality in bacterial populations
\citep{Smith1993}. Traditional tests for deviation from a neutral
coalescent are typically based on single polymorphic sites and often
require information about the number of generations
\citep{Hartl2007}. As such information is not available for clinical
cancer studies, we therefore make a spatial argument about clonal
expansions. We assume that due to the large population sizes of cancer
cells, genetic drift is not significant. In a setting of neutral
evolution where all sequences have essentially the same fitness, sequences
randomly move across the mutational landscape leading to a uniform
distribution of sequences in that space (Figure \ref{fig:ith}A) with
no selective sweeps or clonal expansions. On the contrary, if strong
selective pressure favours specific mutations (Figure \ref{fig:ith}B),
sequences are more likely to survive and be sampled from the favoured
regions leading to local clustering of sequences on the mutational
landscape (Figure \ref{fig:ith}C).

Besag's $L(r)$ \citep{Besag1977}, a variance-stabilised transformation
of Ripley's $K(r)$, \citep{Ripley1977} is a function used in spatial
statistics to test for non-homogeneity, i.e. spatial clustering, of
points in a plane. $\lambda K(r)$ describes the expected number of
additional random points within a distance $r$ of a typical random
point of an underlying Poisson point process with intensity $\lambda$.
The empirical estimate of Ripley's $K$ for $n$ points with pairwise
distances $d_{ij}$ and average density $\hat{\lambda}$ is defined as

\begin{equation*}
  \hat{K}(r) = \frac{1}{\hat{\lambda}n} \sum_{i\ne j} I(d_{ij} < r),
\end{equation*}
where $I$ is the indicator function.
In case of complete spatial randomness (CSR), the expectation of
$K(r)$ is $\pi r^2$. Besag's $L$ is the transformation $L(r) =
\sqrt{K(r)/\pi}$ thereof and is under CSR in expectation linear in
$r$. Therefore plotting $r - \hat{L}(r)$ can be used as a graphical
indication of deviation from CSR. We use a simulation approach to
estimate significance bands for $L(r)$ \citep{Baddeley2005}. 

The
clonal expansion index CE for a dataset (typically samples taken from
a single patient) is then defined as the maximum ratio between the
distance of the observed L-value ($L_{o}(r)$) and the theoretical
L-value under CSR ($L_{t}(r)$) and one-half the width of the two-sided
simulated significance band $C(r)_{u}$ ($u$ for upper significance
band):

\begin{equation}   
  CE = \max_r\left(\frac{|L_o(r) - L_t(r)|}{C_u(r) - L_t(r)}\right)
\end{equation} 

A value of CE $<1$ therefore suggests CSR in the point set,
whereas a CE value $>1$ indicates local spatial clustering. We
conducted coalescence simulations to confirm that the clonal expansion
index distinguishes between trees with normal and elongated branch
lengths between populations (black and red distributions, Figure
\ref{fig:sim}B).

\subsubsection*{Testing for star topology and molecular clock}
Tree reconstruction methods may include positive or negative
assumption of a molecular clock which will significantly influence the
reconstruction accuracy. It is of particular interest in cancer
biology whether evolution is governed by constant or changing rates of
evolutionary change. Furthermore, it is still debated whether disease
progression follows a (structured) tree-like pattern of evolution or
if subpopulations are emitted in radial (star-like) fashion from a
small population of stem-like progenitors (see \citep{Adams2008}).

We implement two tests for tree-likeness and molecular clock in the
\emph{MEDICCquant} package to help answer these questions. We model
genomic events $x$ as generated from a Poisson process with rate
$\rho$. The expected number of events is then linear in time: $E[X] =
\rho t$.  Assuming $\rho=1$, where the process is not time-calibrated,
the observed distance $\hat{X}$ is the maximum likelihood estimate
(MLE) for the time of divergence. Under asymptotic normality of the
MLE we have that $\hat{X} \sim N(X,X)$. Given a star topology we find
optimal branch lengths that minimise the residual sum of squares
between the optimised distances $x_i^{\text{opt}}$ and the measured
pairwise distances $\hat{X_i}$ for branch $i$. Under the null
hypothesis of star-like evolution this sum of squares
\begin{equation*}
RSS_{\text{star}} = \sum_{i=1}^{n(n-1)/2}\frac{\left(x_{i}^{\text{opt}}-\hat{X_i}\right)}{\sqrt{\hat{X_i}}}^2
\end{equation*}
is then $\chi^2$-distributed with $n(n-1)/2-n$ degrees of freedom,
where $n$ is the number of samples studied, i.e. the number of leaves
in the tree (Tim Massingham, pers. communication).

An analogous procedure can be used for testing whether a tree follows a
molecular clock hypothesis, in which it exhibits constant evolutionary rates
along all branches. In this case the distances of all leaf nodes from
the diploid should be the same. We measure the deviation of the
branch lengths from the diploid from their mean ($\mu(\hat{X})$) by
\begin{equation*}
RSS_{\text{clock}} = \sum_{i=1}^{n} \frac{\left(\mu(\hat{X}) - \hat{X_i}\right)}{\sqrt{\hat{X}}}
\end{equation*}

Because branch lengths do not need to be optimised to a specific topology,
and we are only considering distances to the diploid, the distribution
in this case has $n-1$ degrees of freedom.

\subsection*{Progression and heterogeneity in a case of metastatic
  endometrioid adenocarcinoma}
In the following section we demonstrate MEDICC on a case from the
CTCR-OV03 clinical study \citep{Sala2012}.  This case had advanced
endometrioid ovarian carcinoma and was treated with platinum-based
neoadjuvant chemotherapy. After three cycles of chemotherapy the
patient had stable disease based on RECIST assessment, pre- and
post-chemotherapy CT imaging and a $92\%$ reduction of the tumour
response marker CA125. She then underwent interval debulking surgery
but had residual tumour of $>1$cm at completion. After six moths she
progressed with platinum-resistant disease and died one month later.

Out of 20 biopsy samples 18 satisfied quality control for $>50$ tumour
cellularity and array quality. The dataset included 14 omentum
samples, two samples from the vaginal vault (VV) and two samples from
the external surface of the bladder (BL). The BL and VV samples were
taken prior to chemotherapy and the omental samples were collected at
interval-debulking surgery after three cycles of chemotherapy.

All samples were CN profiled with Affymetrix SNP 6.0 arrays and
segmented and compressed using PICNIC \citep{Greenman2010} and
CGHregions \citep{Wiel2007}. Pairwise evolutionary distances between
all samples were estimated with MEDICC. The distance distribution was
tested for the molecular clock hypothesis using MEDICCquant and showed
strong non-clock like behaviour ($p=2.89 x 10^{-14}$, Figure
\ref{fig:example}A). Tree reconstruction was performed by MEDICC using
the Fitch-Margoliash algorithm \citep{Fitch1967}. MEDICCquant detected
a high degree of clonal expansion ($CE = 1.24$) as can be seen in
the strong spatial clustering of samples on the mutational landscape
(Figure \ref{fig:example}B). MEDICC counted a median of 204 genomic
events relative to the diploid and a median of 146 between all
pairwise comparisons. Tree reconstruction showed good support values
for the omental and BL/VV subclades, suggesting strong spatial
heterogeneity. The patient also showed strong temporal heterogeneity,
as there were large evolutionary distances between samples before and
after neoadjuvant chemotherapy (temporal heterogeneity index 3.78,
Figure \ref{fig:example}B). However, temporal and spatial
heterogeneity in this case are indistinguishable because the BL/VV
samples coincide with the biopsy samples, whereas all omentum samples
were taken at surgery.

Ancestral reconstructions using MEDICC showed loss-of-heterozygosity
(LOH) events on chromosome 17q (see internal node profiles in Figure
\ref{fig:example}A) that often coincide with deleterious mutations in
BRCA2 and TP53 \citep{Archibald2012}. The most prominent contributors
to the clonal expansions of the subgroup surrounding sample S01 seemed
to be chromosomal amplifications on chromosomes 6, 8, 11 and 14; as
well as LOH on chromosome 15. We also detected large LOH events on
chromosomes 4, 5, 9, 10, 13, 14, 16 and 17 (Figure
\ref{fig:example}C).

\section*{Conclusions}
While significant progress has been made recently to understand tumour
heterogeneity through extensive multiple sampling studies and
experimental efforts, few algorithms have been developed to target the
specific questions raised by such datasets. MEDICC is our contribution
to better reconstruct the evolutionary histories of cancer within a
patient and propose unbiased quantification of heterogeneity and the
degree of clonal expansion.

We have shown the success of these efforts in simulations and their
utility in the example discussed in this article. Further examples
that also elaborate on the connection between clonal expansion and
heterogeneity and patient outcome can be found in our extensive
clinical study (Schwarz et al 2013, in prep.).

As discussed above we attribute the increase in reconstruction
accuracy mainly to two factors. First, MEDICC makes efficient use of
the available phylogenetic information by phasing parental alleles
using the minimum evolution criterion, which has to our knowledge not
been attempted before. Second, MEDICC models actual genomic events
that change CN and incorporates biological constraints such as loss-of
heterozygosity, which is not the case in breakpoint-based approaches.

The loss of reconstruction accuracy of \emph{TuMult} relative even to
naive approaches using Euclidean distances is most likely due to the
fact that \emph{TuMult} was designed for fewer leaf nodes (typically
around 4, \emph{Letouze, personal communication}). It is worth
stressing that, unlike its competitors, MEDICC is not linked to a
specific data collection platform. Data from SNP arrays can be used,
as well as sequencing-based datasets or any other method that returns
absolute copy numbers.

Future work will focus on reductions of algorithmic complexity as well
as the integration of SNP data into the reconstruction process. Once
this is achieved and sufficient curated training data is available,
the FST approach allows us to extend our approach to a full
probabilistic model of cancer evolution.

\section*{Methods} 
SNP array data for the example from the OV03/04 study can be accessed
at the NCBI Gene Expression Omnibus under accession number
\emph{GSE40546}.

\subsection*{Simulation of tumour evolution}
Coalescent trees were simulated using the \emph{APE} R package
\citep{Paradis2004}. Simulation of genome evolution on these trees was
performed by custom code, released as the \emph{SimCopy} R package
\citep{Sipos2013}. \emph{SimCopy} relies on the \emph{PhyloSim}
package \citep{Sipos2011} in order to perform the simulations on the
level of abstract ``genomic regions''.  The genomic regions are
encoded in a sequence of integers, with the sign representing their
orientation. The package then uses modified \emph{PhyloSim} processes
in order to simulate deletion, duplication, inversion, inverted
duplication and translocation events happening with rates specified by
the user. The number of genomic regions affected by each of these
events is modelled by truncated Geometric+1 distributions.  After
simulating genome evolution, CN profiles are reported for leaf and
internal nodes.  Genomes were simulated using 15 leaf nodes, a root
size of 100 segments and an average event length of 12 segments to
allow for overlapping events. Event rates covered the following set:
$0.02, 0.03, 0.04, 0.05, 0.07, 0.1, 0.13, 0.15, 0.18, 0.2$. Individual
event rates were modified with the following factors: deletions:
$0.3$, duplications: $1.0$, inverted duplications: $0.1$, inversions:
$0.2$, translocations: $0.2$. All parameters were chosen such that the
leaf node CN distributions are similar in shape to CN distributions
from experimental data in the clinical study Schwarz \emph{et al.}
2013 (in prep.).

\subsection*{Implementation of MEDICC}
All FST and FSA algorithms were implemented using OpenFST
\citep{Allauzen2007}. MEDICC was written in Python, while
implementation of time-critical parts used C. For the Fitch-Margoliash
implementations we used the Phylip package \citep{Felsenstein2009}.
MEDICC is available at \url{https://bitbucket.org/rfs/medicc} and has
been tested on Windows and UNIX-based systems.

The quantitative analysis of MEDICC results was done in R and all
necessary functions are implemented in the \emph{MEDICCquant} package
included in the MEDICC distribution. Spatial statistics were computed
using the \emph{spatstat} package \citep{Baddeley2005}, and for kernel
manipulations the \emph{kernlab} package was used
\citep{Karatzoglou2004}.

\section*{Authors contributions}
RFS designed and implemented MEDICC and wrote the ms. AT designed and
implemented simulation comparisons with \emph{TuMult}. BS designed and
implemented the genome simulations. JDB designed the clinical study
and supervised the analysis of clinical data. NG and FM supervised the
phylogenetic and machine learning aspects of the project. NG, FM and
JDB corrected and edited the ms and all authors approved the final
version.

\section*{Acknowledgements}
\ifthenelse{\boolean{publ}}{\small}{} The authors would like to thank
Gonzalo Iglesias and Adria de Gispert from the Cambridge University
Engineering Department for input on the FST implementations. We would
further like to thank Eric Letouze for advice on optimal preparation
of our simulated datasets for \emph{TuMult}.


{\ifthenelse{\boolean{publ}}{\footnotesize}{\small}
 \bibliographystyle{bmc_article}  
  \bibliography{medicc} }     


\ifthenelse{\boolean{publ}}{\end{multicols}}{}


\clearpage
\section*{Figures}
  \subsection*{Figure 1 - Evolutionary CN trees are reconstructed in three steps}
  \begin{figure*}[h]
    \centerline{\includegraphics[width=\textwidth]{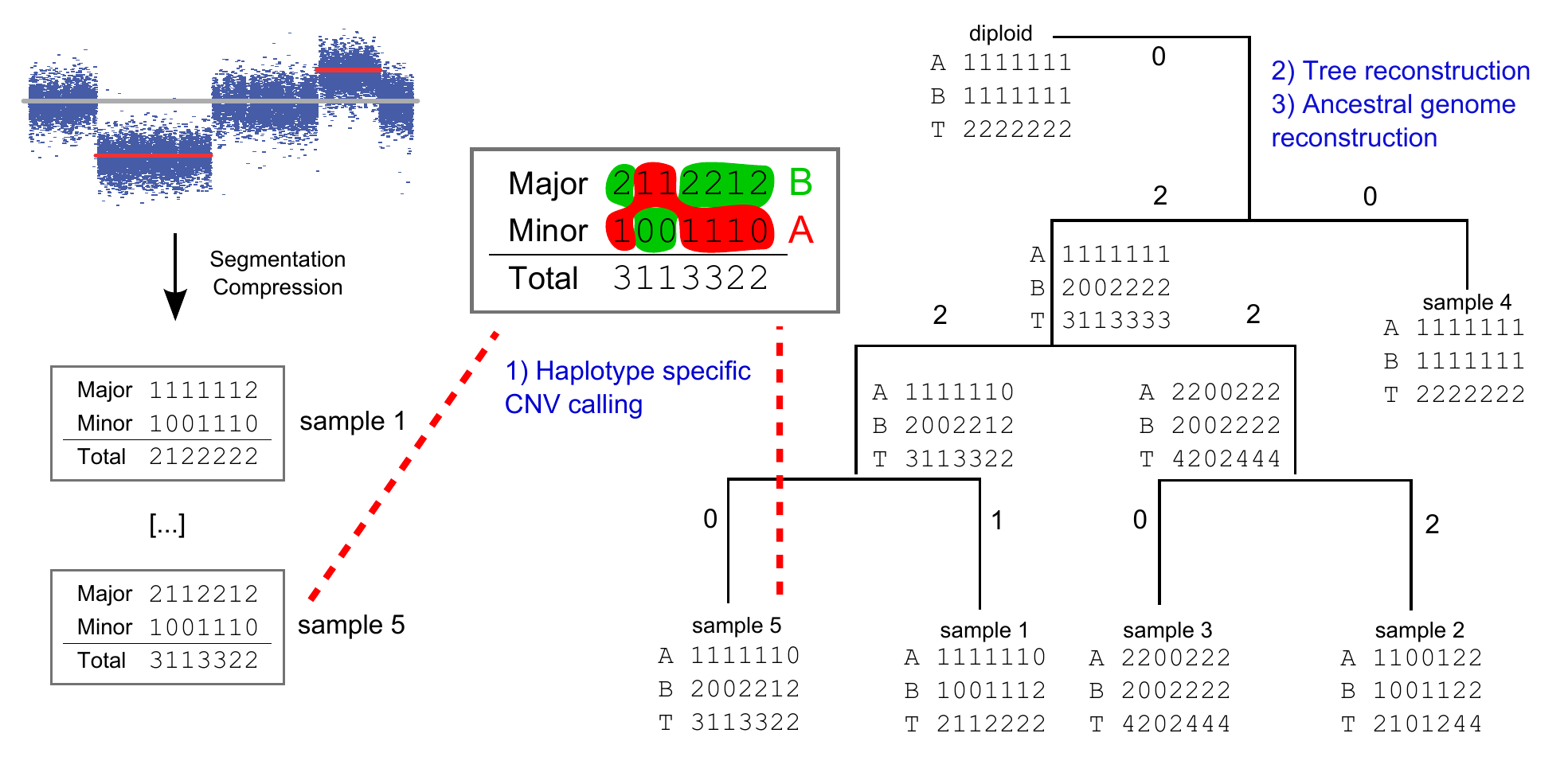}}
    \caption{\label{fig:3steps} \textbf{Evolutionary CN trees are
        reconstructed in three steps:} 1) After segmentation and
      compression, major and minor alleles are phased using the
      minimum event criterion. 2) The tree topology is reconstructed
      from the pairwise distances between genomes. 3) Reconstruction
      of ancestral genomes yields the final branch lengths of the
      tree, which correspond to the number of events between genomes.}
  \end{figure*}

\clearpage
  \subsection*{Figure 2 - Efficient distance calculation is enabled
    via a transducer architecture}
  \begin{figure*}[h]
    \centerline{\includegraphics[width=\textwidth]{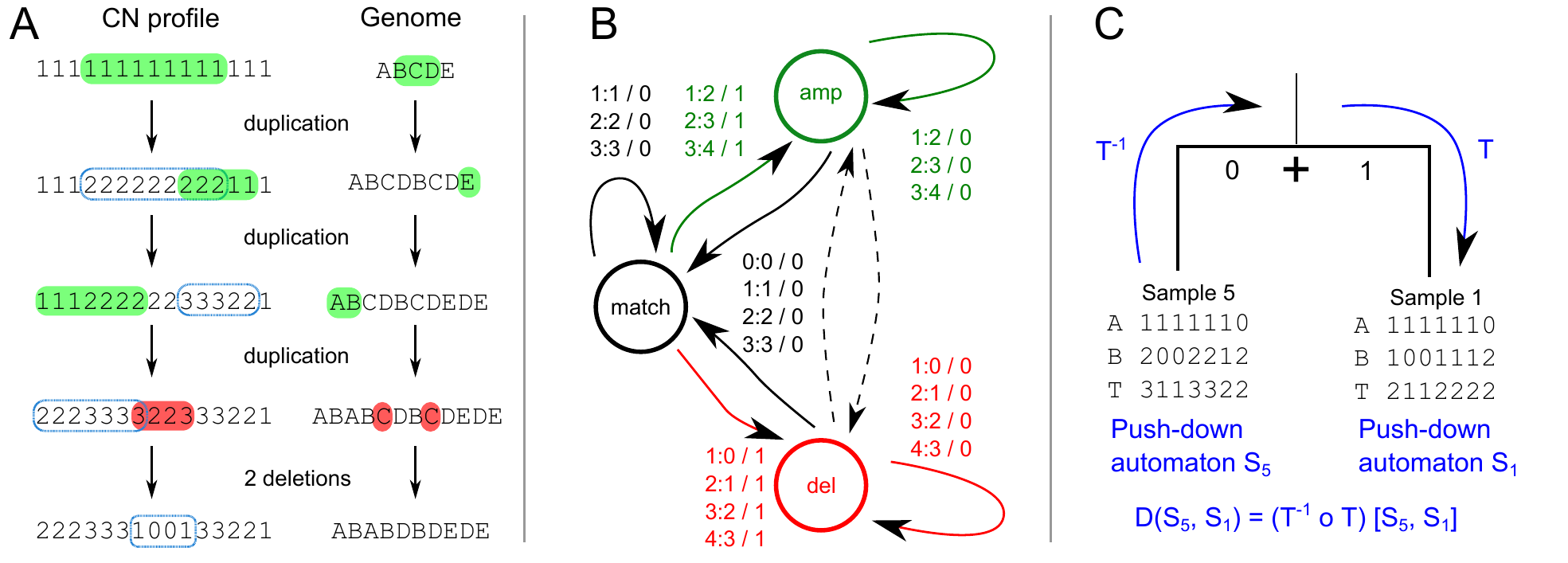}}
    \caption{\label{fig:fst} \textbf{Efficient distance calculation is
        enabled via a transducer architecture:} A) Overlapping genomic
      rearrangements modify the associated CN profiles in different
      ways. Amplifications are indicated in green, deletions in
      red. The blue rectangles indicate the previous event. B) The
      one-step minimum event transducer describes all possible edit
      operations achievable in one event. This FST is composed $n$
      times with itself to create the the full minimum event FST
      $T$. Edge labels consist of an input symbol, a colon and the
      corresponding output symbol, followed by a slash and the weight
      associated with taking that transition. C) The minimum event FST
      $T$ is asymmetric and describes the evolution of a genomic
      profile from its ancestor. Composed with its inverse this yields
      the symmetric minimum event distance $D$.}
  \end{figure*}
 
 \clearpage 
  \subsection*{Figure 3 - Parental alleles are phased using
    context-free grammars}

  \begin{figure*}[h]
    \centerline{\includegraphics[width=\textwidth]{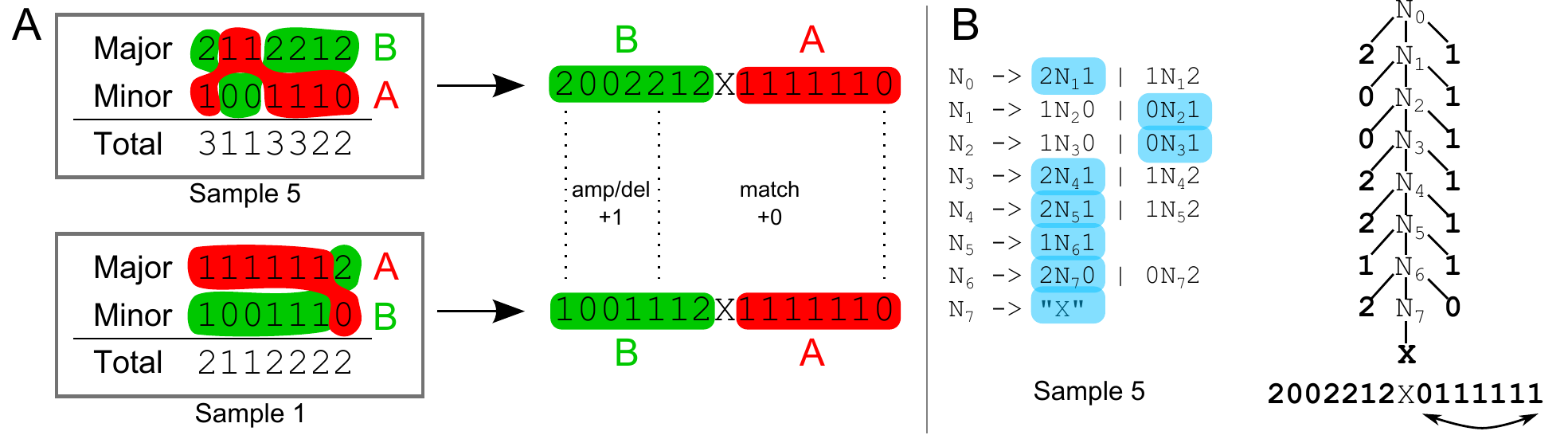}}
    \caption{\label{fig:cfg} \textbf{Parental alleles are phased using
        context-free grammars:} A) Allelic phasing is achieved by
      choosing consecutive segments from either the major or minor
      allele which minimise the pairwise distance between profiles. B)
      The set of all possible phasing choices is modelled by a
      context-free grammar. In this representation, the order of the
      regions' CN values on the second allele is reversed, in order to
      match the inside-out parsing scheme of CFGs. That way every
      possible parse tree of the grammar describes one possible
      phasing.}
  \end{figure*}
 
 \clearpage 
  \subsection*{Figure 4 - MEDICC improves reconstruction accuracy over
  competing methods}
  \begin{figure*}[h]
    \centerline{\includegraphics{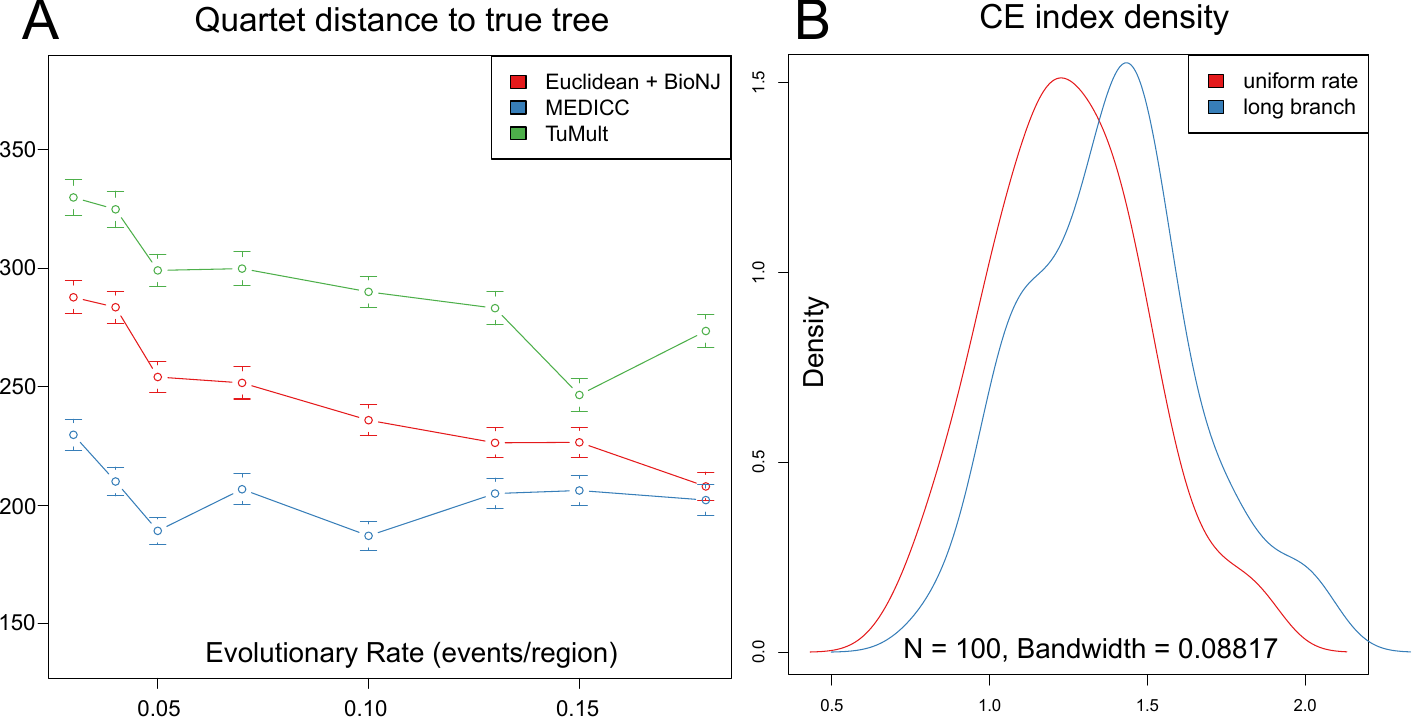}}
    \caption{\label{fig:sim} \textbf{MEDICC improves reconstruction
        accuracy over competing methods:} A) Simulations results show
      the improvement of reconstruction accuracy for MEDICC over naive
      methods (BioNJ clustering on Euclidean distances between CN
      profiles, red) and competing algorithms (TuMult, green). B)
      Density estimates of clonal expansion indices for neutrally
      evolving trees (red) and trees with induced long branches as
      created by clonal expansion processes (blue) show the ability of
      MEDICC to detect clonal expansion.}
  \end{figure*}

\clearpage
\subsection*{Figure 5 - MEDICC quantifies ITH from the locations of
  genomes on the mutational landscape}
\begin{figure*}[h]
  \centerline{\includegraphics{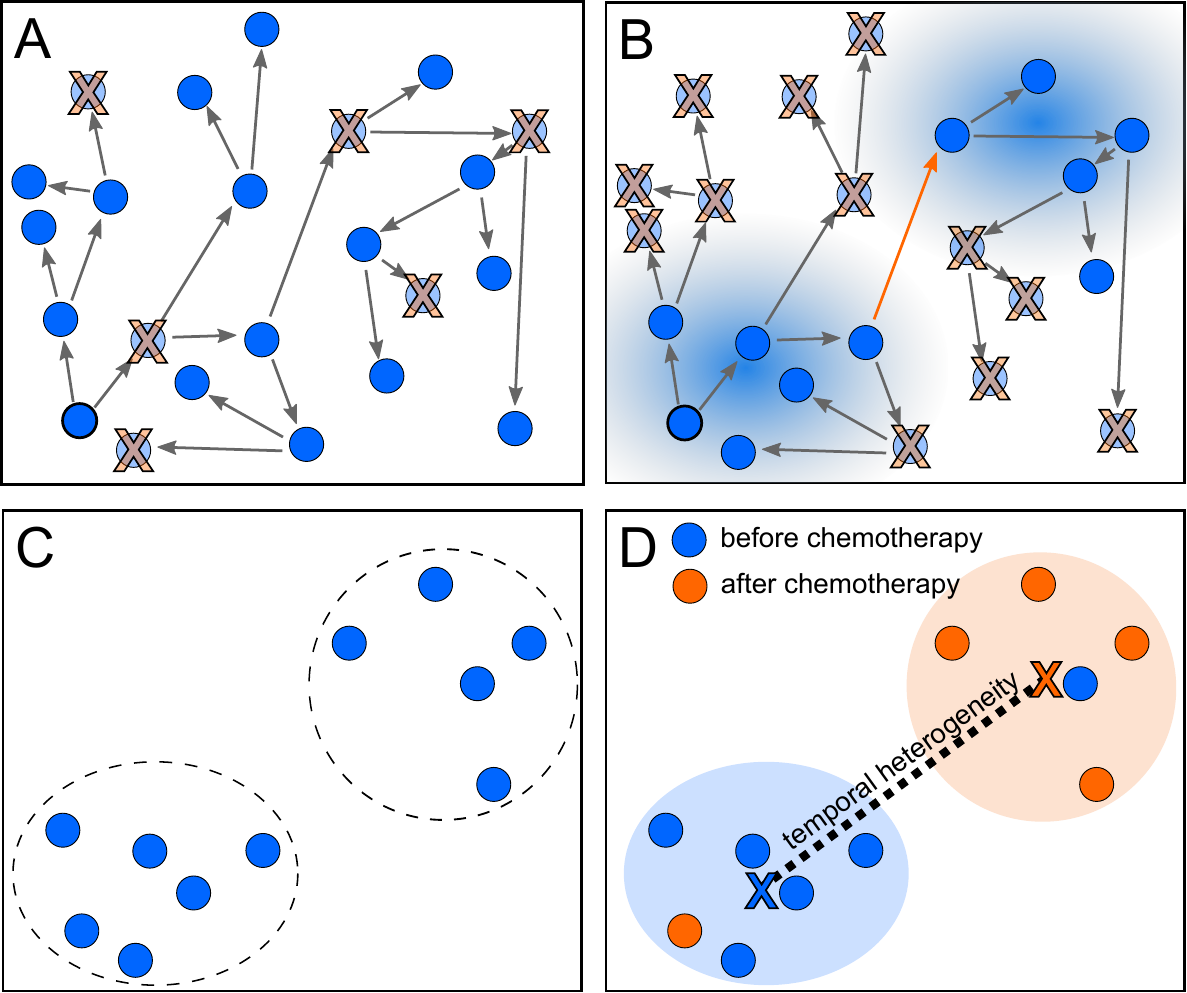}}
  \caption{\label{fig:ith} \textbf{MEDICC quantifies ITH from the
      locations of genomes on the mutational landscape:} A) If no or a
    homogeneous selection pressure is applied, cells proliferate and
    die randomly across the mutational landscape, leaving the
    surviving cells spatially unclustered. B) If the fitness landscape
    favours specific mutations (blue shaded areas), genomes inside
    those areas are more likely to survive, those outside more likely
    to die. The ability of a tumour for a clonal expansion into
    distant fitness pockets depends on its mutation potential per
    generation (long orange arrow). This leads to C) a situation where
    distinct subpopulations/clonal expansions are present in a tumour,
    indicating a generally high potential for a tumour to adapt to
    changing environments. D) The mutational landscape additionally
    allows estimates of average distance between two subgroups of
    samples, here before (blue) and after (orange) chemotherapy. The
    distance between the two subgroups is defined as the distance of
    the robust centres of mass (blue and orange X). This robust centre
    of mass is computed omitting the single most distant point of each
    subgroup (blue and orange samples in the orange and blue subgroups
    respectively), making the statistic more resistant towards
    outliers.}
\end{figure*}
\clearpage
  \subsection*{Figure 6 - Application to a case of endometrioid cancer}
  \begin{figure*}[h]
    \centerline{\includegraphics[width=\textwidth]{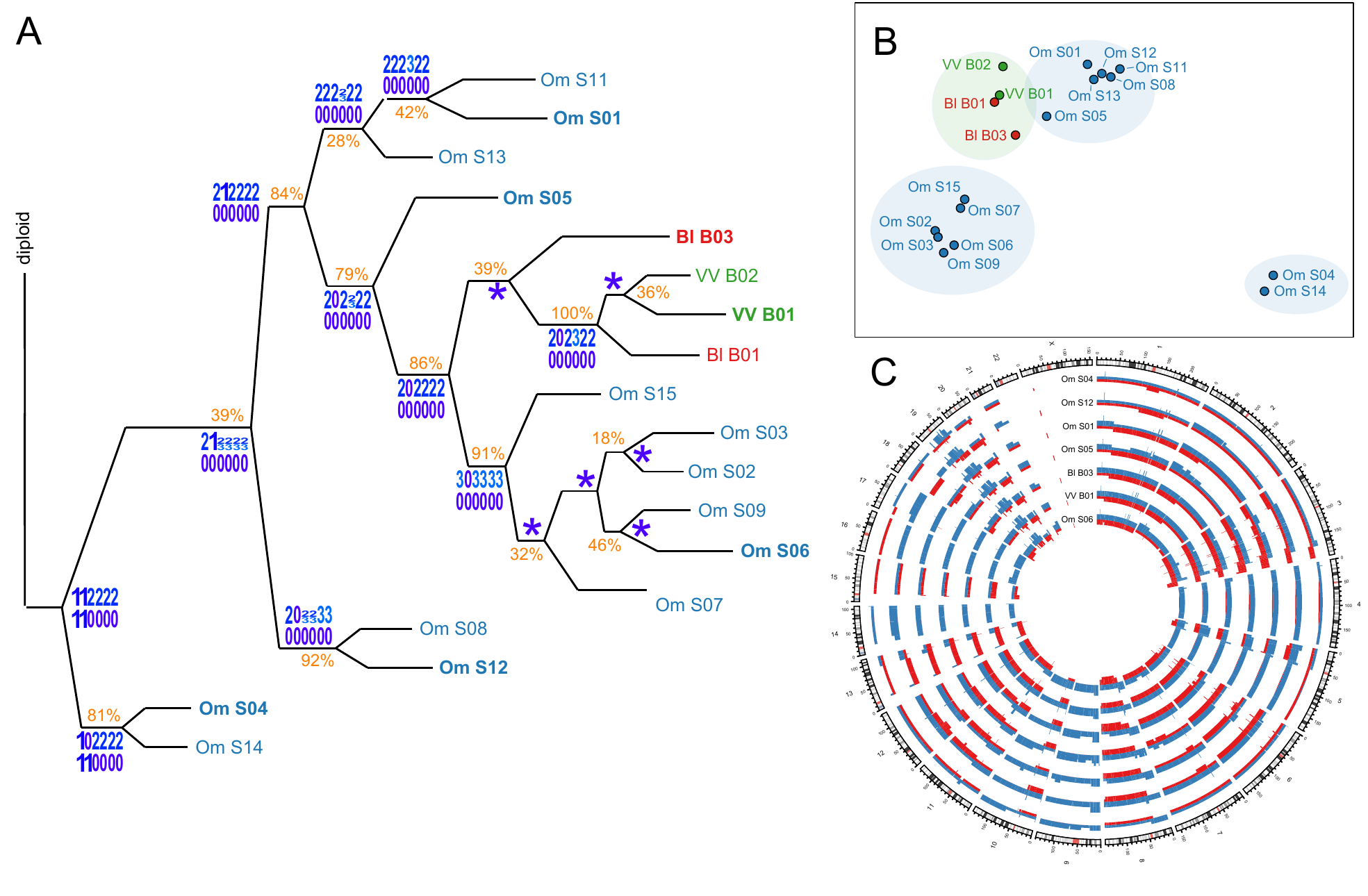}}
    \caption{\label{fig:example} \textbf{Application to a case of
        endometrioid cancer:}  A) Evolutionary tree of the OV03-04 case
      reconstructed from whole genome CN profiles. Approximate support
      values indicate how often each split was observed in trees
      reconstructed after resampling of the distance matrix with added
      truncated Gaussian noise. MEDICC performs reconstruction of
      ancestral CN profiles. Here, the (compressed) ancestral profiles
      for chromosome 17 are given as an example and MEDICC depicts
      unresolved ambiguities in the form of sequence logos. A star
      indicates no change compared to its ancestor. B) Ordination of
      the samples using kPCA shows four clear clonal expansions,
      comprising three separate Omentum groups and the Bl/VV group. C)
      Circos plot of selected genomic profiles (marked in bold in the
      tree) shows the extent of chromosomal aberrations across the
      genome. The two phased parental alleles are indicated in red and
      blue.}
  \end{figure*}

\clearpage    






\end{bmcformat}
\end{document}